\documentclass{webofc}

\usepackage[varg]{txfonts}  
\usepackage{hyperref}
\usepackage{url}
\usepackage{siunitx}
\DeclareSIUnit{\c}{c}
\hypersetup{colorlinks=true,citecolor=blue,urlcolor=blue,linkcolor=blue}

\newcommand{\vb}{\vec}
\renewcommand{\vec}[1]{\mathrm{\mathbf{#1}}}
\newcommand{\dd}[2][]{\mathrm d^{#1}{#2}\,} 
\newcommand{\dv}[2][]{\frac{\dd{#1}}{\dd{#2}}}
\newcommand{\pdv}[2][]{\frac{\partial{#1}}{\partial{#2}}}

\newcommand{\Conetwo}{\mathcal {C}^{1\leftrightarrow 2}}
\newcommand{\Ctwotwo}{\mathcal{ C}^{2\leftrightarrow 2}}

\begin{document}
\title{Jet momentum broadening beyond the jet quenching parameter from QCD kinetic theory }

\author{\firstname{Alois} \lastname{Altenburger}\inst{1}\fnsep
    \and
        \firstname{Kirill} \lastname{Boguslavski}\inst{1}\fnsep
             \and
        \firstname{Florian} \lastname{Lindenbauer}\inst{1}\fnsep\thanks{Speaker, \email{florian.lindenbauer@tuwien.ac.at}, \\
        FL is a recipient of a DOC Fellowship of the Austrian Academy of Sciences at TU Wien (project 27203) and is supported by the Austrian Science Fund (FWF) under Grant DOI 10.55776/P34455. The results have been achieved in part using the Vienna Scientific Cluster (VSC), project 71444.
}
}

\institute{Institute for Theoretical Physics, TU Wien, Wiedner Hauptstraße 8-10, 1040 Vienna, Austria 
          }

\abstract{Using QCD kinetic theory, we study momentum broadening of jets using the full broadening probability (elastic collision kernel) $C(\vb q_\perp)$ extracted during simulations of the nonequilibrium initial stages in heavy-ion collisions. We find that small momentum transfer is more likely than in thermal equilibrium, particularly along the beam axis. Our extraction represents a significant step towards a more realistic modeling of jet quenching during the initial stages.
}
\maketitle
\section{Introduction}
\label{intro}
The suppression of high-energy hadrons in heavy-ion collisions is considered to be one of the signatures of the quark-gluon plasma created therein. In particular, the energy loss of energetic hadrons in many models is tightly connected to momentum broadening of a parton moving through the plasma.

Recently, much emphasis has also been put on modeling jet quenching during the initial stages \cite{Andres:2019eus, Ipp:2020nfu, Carrington:2022bnv, Avramescu:2023qvv, Boguslavski:2023alu, Barata:2024xwy}, which are still far from equilibrium, and where a hydrodynamic description of the system is less reliable.
In the commonly accepted weak-coupling picture, immediately after the collision, the system is dominated by large gluonic fields (glasma), which can then be modeled as quasi-particles using kinetic theory, where the system isotropizes and a hydrodynamic description becomes applicable \cite{Berges:2020fwq}.

\section{Modeling jet energy loss}
In the simplest case, jet energy loss is described in the so-called harmonic approximation, where the dipole cross section $C(\vb b)$ is approximated by its small $b$ limit,
\begin{align}
    C(\vb b) = \int\frac{\dd[2]{\vb q_\perp}}{(2\pi)^2}\left(1-e^{i\vb q_\perp\cdot \vb b}\right)C(\vb q_\perp),\label{eq:fouriertrafo} && C(\vb b)\approx \frac{1}{4}\hat q\vb b^2, &&     \hat q = \dv[\langle p_\perp^2\rangle]{t}=\int\frac{\dd[2]{\vb q_\perp}}{(2\pi)^2}\vb q_\perp^2 C(\vb q_\perp).
\end{align}
The coefficient $\hat q$ is called the jet quenching parameter. Physically, it represents the amount of transverse momentum broadening per time, and can be obtained via the second moment of the collision kernel $C(\vb q_\perp)$ in momentum space.

\begin{figure}
\centering
\centerline{
\includegraphics[width=0.33\linewidth]{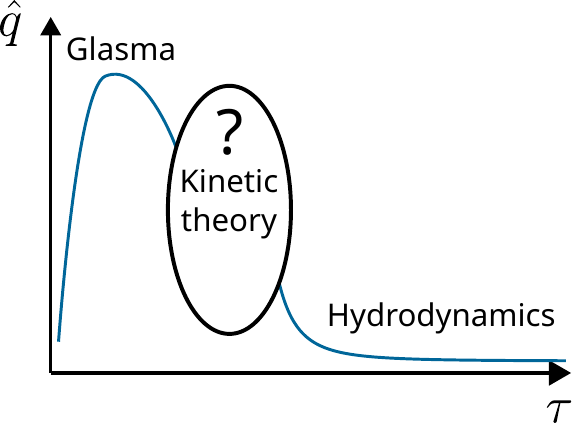}
\includegraphics[width=0.33\linewidth]{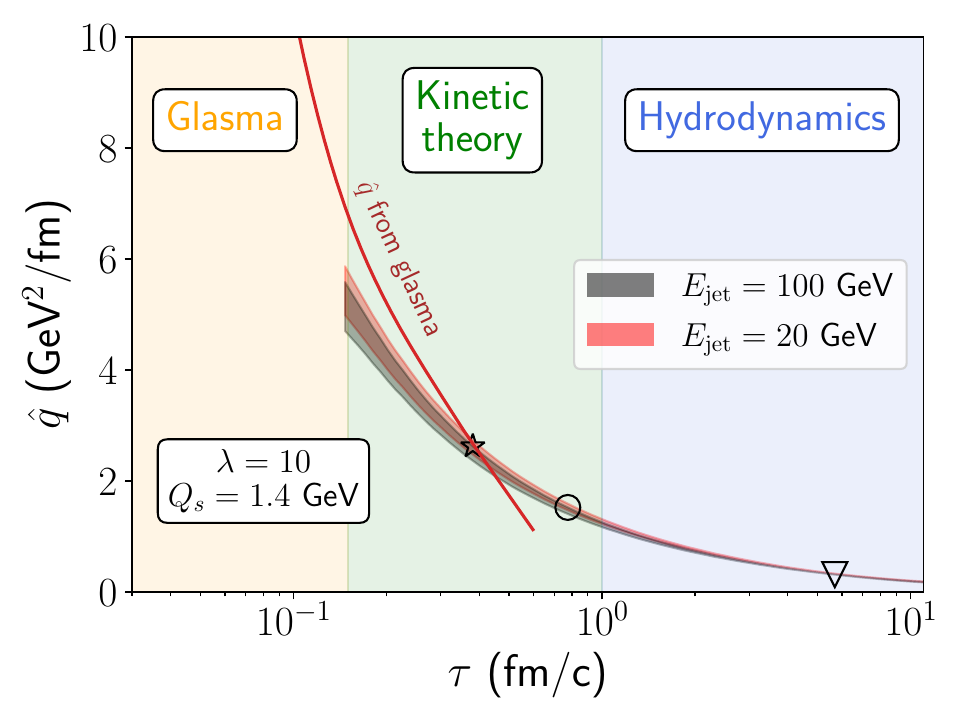}
\includegraphics[width=0.33\linewidth]{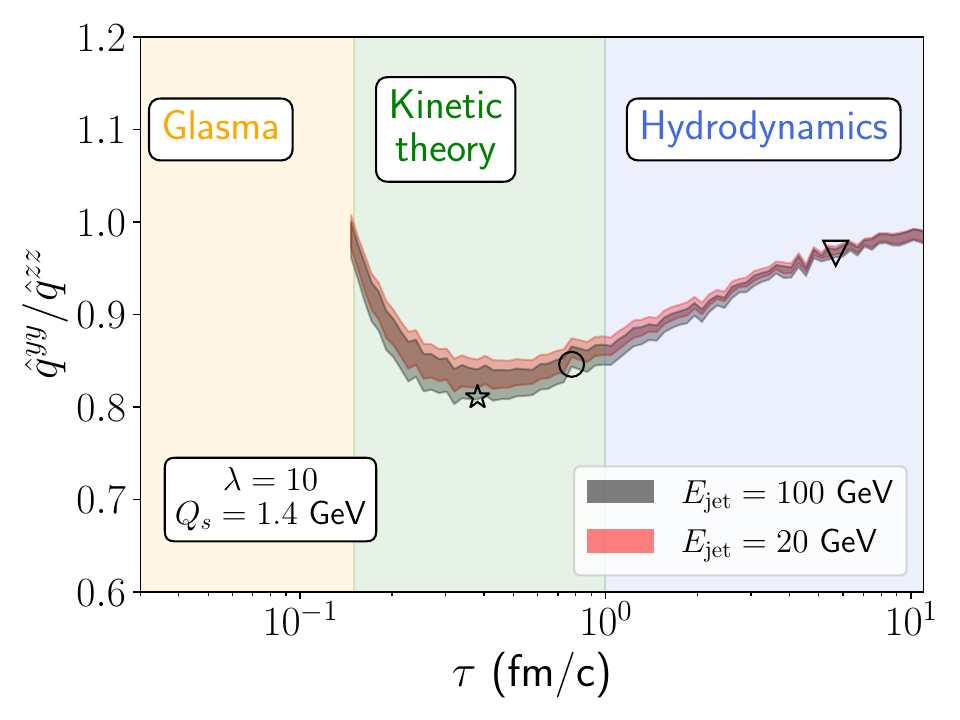}
}
\caption{(Left): Schematic evolution of the jet quenching parameter $\hat q$ during the initial stages in heavy-ion collisions. (Right): Jet quenching parameter obtained from a kinetic theory simulation. Additionally shown is the Glasma result of Ref.~\cite{Ipp:2020nfu}.
Figures taken from Ref.~\cite{Boguslavski:2023alu}.}
\label{fig:qhat-schematic}      
\end{figure}

Within the harmonic approximation \eqref{eq:fouriertrafo}, the gluon emission rate can be obtained analytically \cite{Zakharov:1996fv, Baier:1998yf}, and these results can be used in a phenomenological setup to compare with experimental data.
For instance, in Ref.~\cite{Andres:2019eus}, it was reported that the comparison with experimental data favors $\hat q=0$ for $\tau < \SI{0.6}{\femto\meter/\c}$. However, extractions of this parameter in the Glasma \cite{Ipp:2020nfu, Carrington:2022bnv, Avramescu:2023qvv} 
show that this parameter is actually very large. In these proceedings, we recapitulate how $\hat q$ has been obtained in the kinetic theory stage in Ref.~\cite{Boguslavski:2023alu}, connecting the Glasma and hydrodynamic regimes, 
as sketched in Fig.~\ref{fig:qhat-schematic}. We will then proceed in section \ref{sec:beyondqhat} by generalizing this extraction and obtaining the elastic collision kernel $C(\vb q_\perp)$.

\section{QCD kinetic theory, initial conditions and time markers}
Weakly-coupled QCD admits an effective kinetic description \cite{Arnold:2002zm}, where the particle distribution function $f(\vb p, t)$ is evolved in time using the Boltzmann equation and an initial condition
\begin{align}
\pdv[f(\vb p)]{\tau}- \frac{p_z}{\tau} \pdv[f(\vb p)]{p_z} =-\Conetwo[f(\vb p)]- \Ctwotwo[f(\vb p)], && f(\tau_0)=\frac{2A(\xi)\langle p_T\rangle}{\lambda \, p_\xi}\exp\left({-\frac{2p_\xi^2}{3\langle p_T\rangle^2}}\right).
\label{eq:boltzmann_equation}
\end{align}
For simplicity, we consider a system of pure gluons, which are the dominant degree of freedom at early times \cite{Kurkela:2018xxd}. The collision terms describe elastic ($\Ctwotwo$) and inelastic ($\Conetwo$) collisions, where the latter one interpolates between the Bethe-Heitler and LPM limit \cite{Arnold:2002ja}.

We use an implementation based on \cite{kurkela_2023_10409474}, with the soft-gluon exchange regulated by Debye-like screening prescription \cite{AbraaoYork:2014hbk} (for the results of $\hat q(t)$ as obtained in Ref.~\cite{Boguslavski:2023alu}) or regulated by the isotropic hard-thermal loop self-energy \cite{Boguslavski:2024kbd} (for the results of $C(\vb q_\perp)$ in Ref.~\cite{Altenburger:2025}). To initialize the distribution in Eq.~\eqref{eq:boltzmann_equation}, we use the Kurkela-Zhu initial conditions \cite{Kurkela:2015qoa} inspired by the Glasma, 
with $\langle p_T\rangle = 1.8 Q_s$, $p_\xi=\sqrt{p_\perp^2+(\xi p_z)^2}$, $\xi \in \{4,10\}$, $A(\xi=10) =  5.24171$.
Note that this initial condition is overoccupied ($\sim 1/\lambda$), and $\lambda=g^2N_C$ is the 't Hooft coupling.

\begin{figure}
    \centering
    \centerline{
    \includegraphics[width=0.35\linewidth]{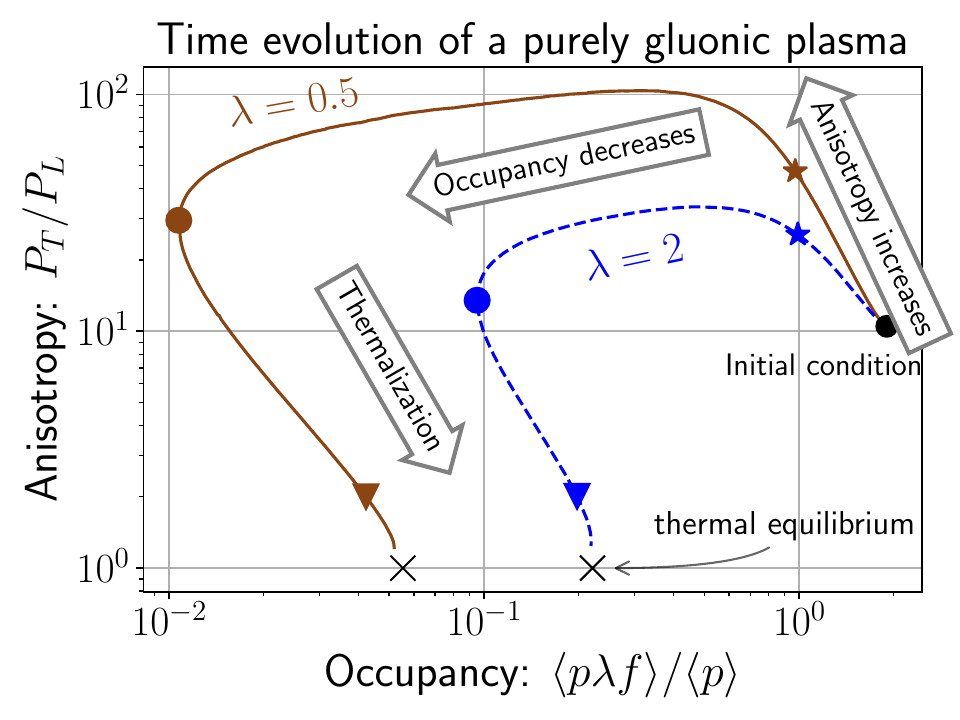}
    \qquad\qquad
    \includegraphics[width=0.33\linewidth]{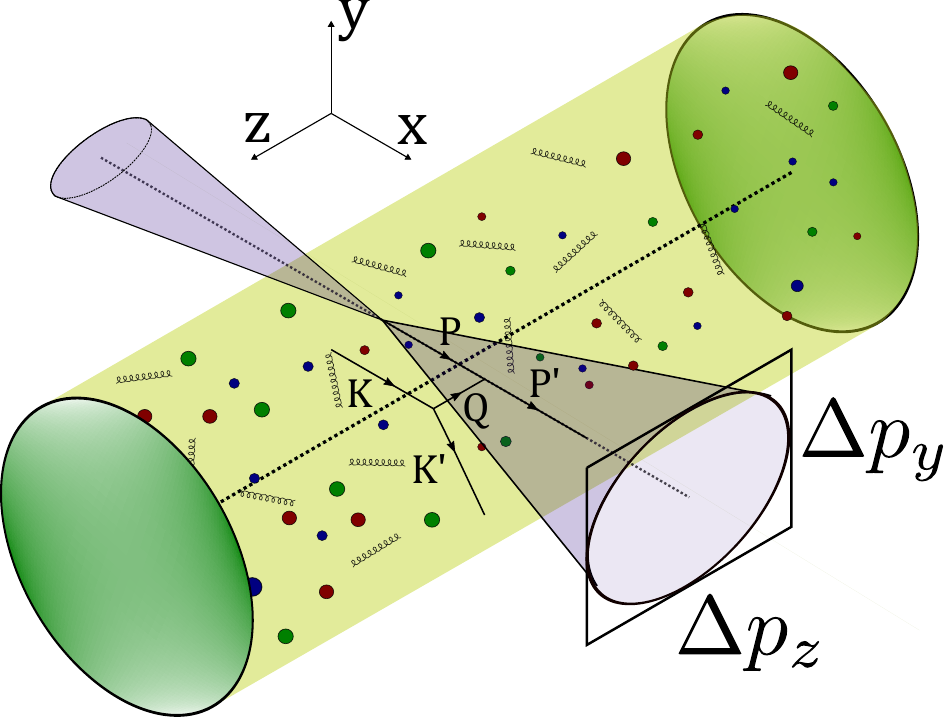}
    }
    \caption{(Left): Time evolution of an initially overoccupied gluonic plasma undergoing longitudinal expansion. The three stages of bottom-up thermalization \cite{Baier:2000sb} are visible and denoted by arrows in the figure. Adapted from \cite{Boguslavski:2023alu}, based on a plot in \cite{Kurkela:2015qoa}. (Right): Sketch of a jet moving in the $x-$ direction through the quark-gluon plasma generated in heavy-ion collisions. The jet accumulates a transverse momentum $\Delta p_z$ and $\Delta p_y$ in the beam- and transverse directions. Figure taken from \cite{Boguslavski:2023waw}.}
    \label{fig:overview_curves}
\end{figure}
Using these initial conditions, we numerically solve the Boltzmann equation \eqref{eq:boltzmann_equation} and follow the system's evolution towards equilibrium. At weak couplings, and for initially overoccupied systems undergoing Bjorken expansion, the plasma follows the bottom-up thermalization picture \cite{Baier:2000sb}, which consists of several stages denoted by arrows in the left panel of Fig.~\ref{fig:overview_curves}. We also introduced the time markers in \cite{Boguslavski:2023alu}, which should guide the eye and roughly separate the three stages in this thermalization picture: The star marker is placed where the occupancy $\langle p f\rangle/\langle p\rangle=1/\lambda$, the circle marker at minimum occupancy and the triangle marker where the pressure ratio $P_T/P_L$ drops below $2$, which indicates that the system has almost isotropized. These markers will be added in subsequent figures in these proceedings.

\section{Results for the jet quenching parameter}
Given a gluon distribution function $f(\vb p,t)$ obtained from numerically solving \eqref{eq:boltzmann_equation},
we may obtain the jet quenching parameter and collision kernel via \cite{Boguslavski:2023alu, Boguslavski:2023waw}
\begin{align}
    \hat q^{ii}=\int_{\substack{q_\perp <\Lambda_\perp\\ p\to \infty}}\dd{\Gamma}\left(q^i\right)^2\left|\mathcal M\right|^2 f(\vb k)\left(1+f(\vb k')\right), &&C(\vb q_\perp)=\int_{\substack{p\to \infty}}\dd{\tilde \Gamma}\left|\mathcal M\right|^2 f(\vb k)\left(1+f(\vb k')\right),
    \label{eq:formula-qhat-collisionkernel}
\end{align}
where we generalize $\hat q=\hat q^{yy}+\hat q^{zz}$ to accommodate directional broadening. In $|\mathcal M|^2$, we screen soft-gluon exchanges using hard-thermal loop resummed propagators \cite{Boguslavski:2023waw}. 
As sketched in Fig.~\ref{fig:overview_curves}, we take $z$ to be the beam axis and $x$ to be the jet direction.
\begin{figure}
    \centering
    \centerline{
    \includegraphics[width=0.33\linewidth, trim={ 0 0.4cm 0cm 0.35cm}, clip]{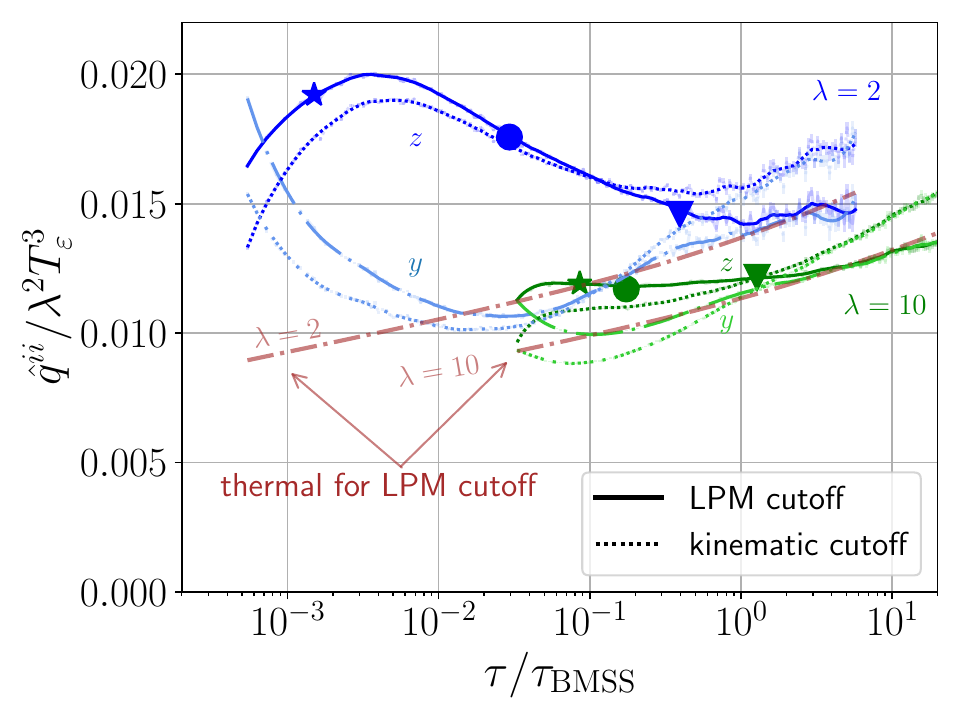}
    \includegraphics[width=0.33\linewidth, trim={ 0 0.4cm 0cm 0.35cm}, clip]{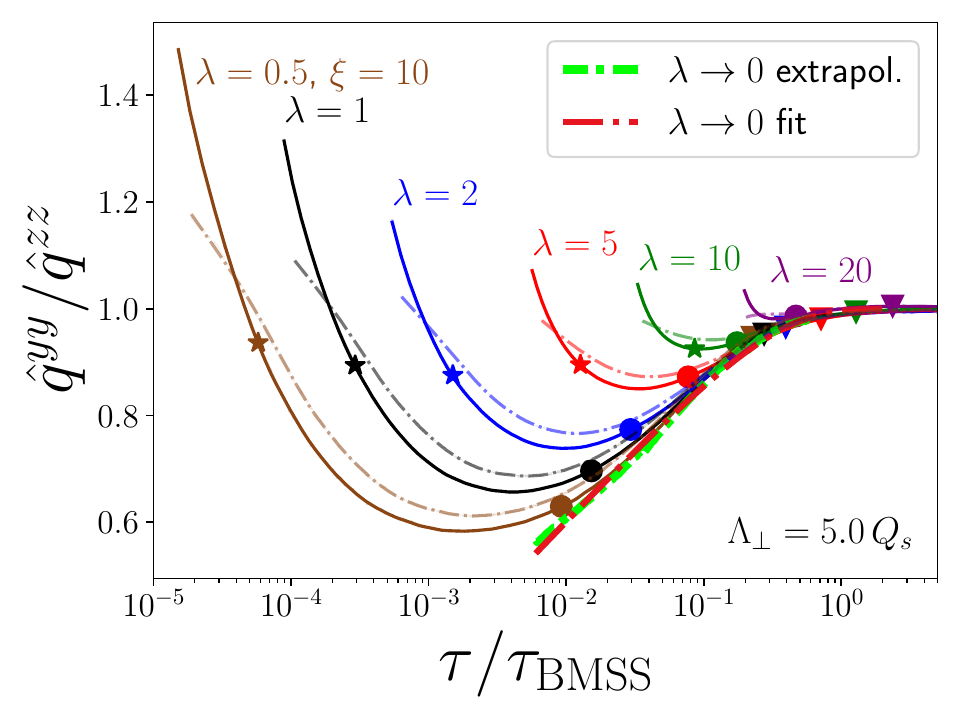}
    \includegraphics[width=0.33\linewidth, trim={ 0 0.4cm 0cm 0.35cm}, clip]{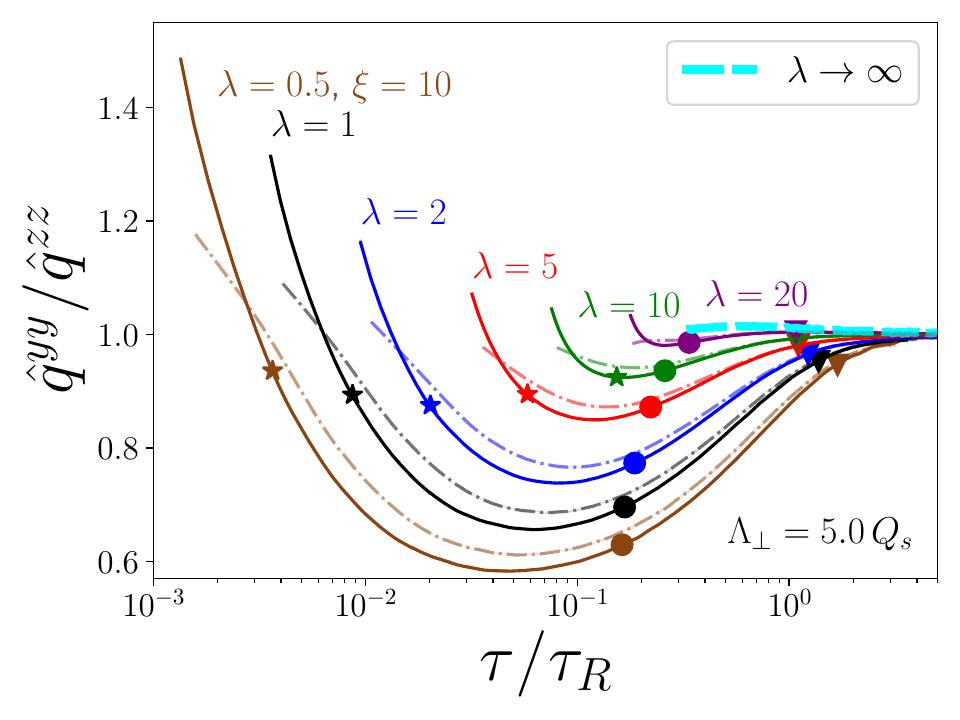}
    }
    \caption{(Left): Time evolution of the jet quenching parameter in a Bjorken expanding plasma for two different couplings (color-coded) and different broadening directions. Comparison of the two cutoff models \eqref{eq:cutoff-models}. Taken from \cite{Boguslavski:2023alu}. (Center and right): The ratio of jet quenching parameters in different directions for different couplings (color-coded). Time is rescaled with $\tau_{\mathrm{BMSS}}$ and $\tau_R$, respectively. Shown are also extrapolations to zero and infinite coupling. Figures taken from \cite{Boguslavski:2023jvg}.}
    \label{fig:qhat-timeevolution}
\end{figure}

Equation \eqref{eq:formula-qhat-collisionkernel}
for $\hat q$ must be supplemented with a model of how the cutoff $\Lambda_\perp$ depends on jet energy $E$ and medium temperature $T_\varepsilon$.
It can be shown that $\hat q$ exhibits a logarithmic dependence on the cutoff $\Lambda_\perp$ for sufficiently large $\Lambda_\perp$ \cite{Boguslavski:2023waw}. 
In Ref.~\cite{Boguslavski:2023alu}, two different models 
\begin{align}
    \Lambda_\perp^{\mathrm{LPM}}(E,T_\varepsilon) \sim g\times (ET_\varepsilon^3)^{1/4}, && \Lambda_\perp^{\mathrm{kin}}(E,T_\varepsilon)\sim (ET_\varepsilon)^{1/2},\label{eq:cutoff-models}
\end{align}
are used, with the proportionality constant fixed to obtain a reference value of $\hat q$ at a given jet energy and temperature \cite{JETSCAPE:2021ehl}. 

In the left panel of Fig.~\ref{fig:qhat-timeevolution}, we show the time evolution of $\hat q^{ii}(t)$ for different couplings and cutoff models \eqref{eq:cutoff-models}. The jet quenching parameter is plotted as a dimensionless ratio over the Landau-matched plasma temperature $T_{\varepsilon}$ (the temperature of a thermal system that reproduces the energy density of the nonequilibrium system). We observe that (apart from very early times) momentum broadening is enhanced in the direction of the beam axis ($z$) than transverse to it ($y$), which is similar to other quasiparticle calculations \cite{Romatschke:2006bb}. Furthermore, the broadening coefficients in different directions differ more than choices of the cutoff model in Eq.~\eqref{eq:cutoff-models}. We also show in this figure the corresponding $\hat q$ obtained in a thermal system with the LPM cutoff, and observe that the nonequilibrium $\hat q$ lies in the same ballpark as
its equilibrium value when compared at the same energy density.\footnote{A similar result has been obtained for the heavy-quark diffusion coefficient \cite{Boguslavski:2023fdm}. }

With $Q_s=\SI{1.4}{\giga\electronvolt}$ \cite{Boguslavski:2023alu, Keegan:2016cpi}, the value of the jet quenching parameter $\hat q(t)$ at initial time is found to be comparable to those reported from the Glasma calculation in Ref.~\cite{Ipp:2020nfu}. These results are shown in the central panel of Fig.~\ref{fig:qhat-schematic}. In the right panel of the same figure, we show the anisotropy ratio in the jet quenching parameter, which for the chosen (and displayed) parameter values reaches a maximum of about $20\%$.

Finally, results of kinetic theory simulations are often presented in time ratios rescaled with a relaxation time $\tau_R=\frac{4\pi\eta/s}{T_\varepsilon(t)}$, in which the first-order hydrodynamic transport coefficient $\eta$ enters. In contrast, parametric estimates of the bottom-up equilibration process \cite{Baier:2000sb} yield a parametrically different equilibration time scale $\tau_{\mathrm{BMSS}}=\left(\frac{\lambda}{4\pi N_\mathrm{C}}\right)^{-13/5}/Q_s$.

Remarkably, rescaling time with $\tau_{\mathrm{BMSS}}$ leads to a collapse of the curves of the ratios of the jet quenching parameters $\hat q^{yy}/\hat q^{zz}$, whereas the rescaling with $\tau_R$ seems to be less useful for this ratio. The results are presented in the central and right panel of Fig.~\ref{fig:qhat-timeevolution}, where also the extrapolation to zero and infinite coupling is performed. These extrapolated curves are interpreted as \emph{limiting attractors} in Ref.~\cite{Boguslavski:2023jvg}.

\section{Going beyond the jet quenching parameter\label{sec:beyondqhat}}
\begin{figure}
    \centering
        \includegraphics[width=0.35\linewidth, trim={ 0 0.4cm 0cm 0.5cm}, clip]{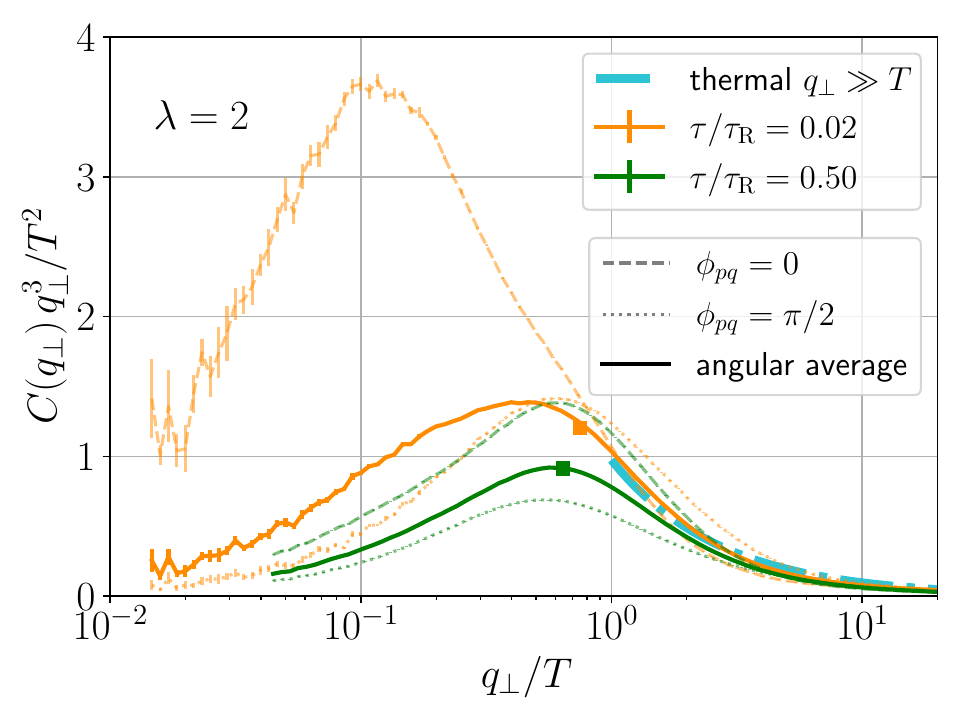}
        \qquad
         \includegraphics[width=0.35\linewidth, trim={ 0 0.4cm 0cm 0.45cm}, clip]{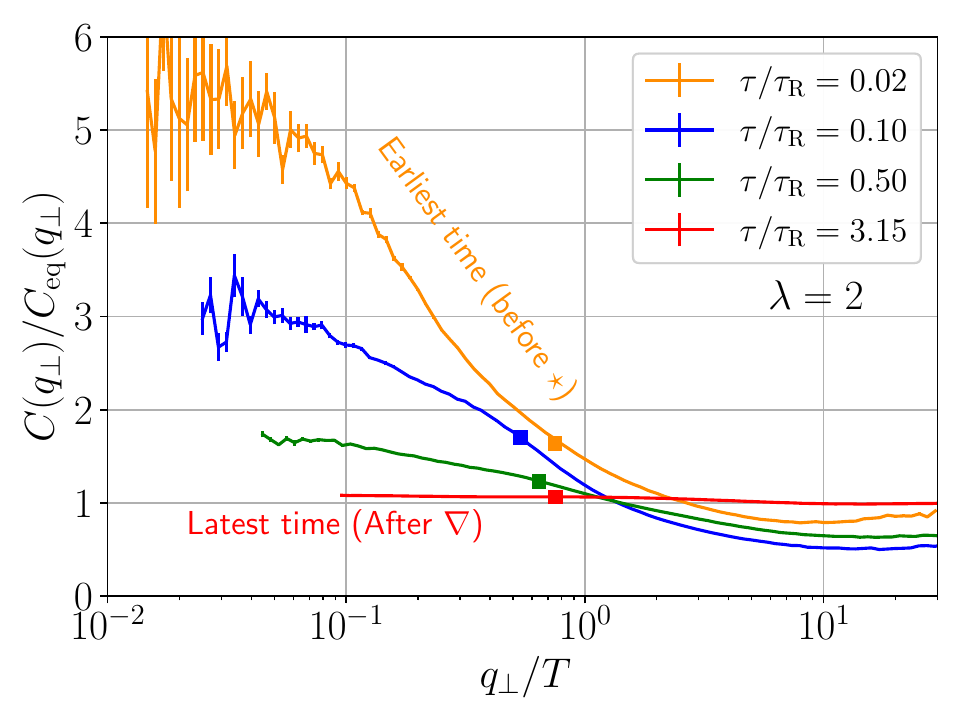}
    \caption{Collision kernel $C(\vb q_\perp)$ for different times for $\lambda=2$. (Left): Integrand for the jet quenching parameter $\hat q$ for
    different angles of $\vb q_\perp$ and the angular average. Squares represent the Debye mass.
    (Right): Normalized to the Landau-matched thermal equilibrium kernel. Figures taken from \cite{Altenburger:2025}.}
    \label{fig:collision-kernel}
\end{figure}
There are several reasons to go beyond the simple harmonic approximation $C(\vb b)\approx \frac{1}{4}\hat q\vb b^2$ in Eq.~\eqref{eq:fouriertrafo}. For example, recently, formalisms have been developed to obtain the spectrum or rate for the full collision kernel (see, e.g., \cite{Caron-Huot:2010qjx, Andres:2020vxs, Moore:2021jwe}). Additionally, 
the jet quenching parameter as formulated in Eqs.~\eqref{eq:fouriertrafo} or \eqref{eq:formula-qhat-collisionkernel} for $\hat q$ has an inconvenient dependence on a UV cutoff.
The collision kernel $C(\vb q_\perp)$ does not suffer from this cutoff dependence.
Similarly as in the case of the jet quenching parameter $\hat q$, we obtain the collision kernel using Eq.~\eqref{eq:formula-qhat-collisionkernel}.

We present our results for the collision kernel in Fig.~\ref{fig:collision-kernel}. In the left panel, we plot the integrand needed for the evaluation of $\hat q$ (see Eq.~\eqref{eq:fouriertrafo}) and show two distinct times of the evolution. For each time, we show two angles ($\phi_{pq}=0$ means broadening along the beam axis) and the angular averaged kernel. Additionally, we show the analytic thermal result for large $q_\perp$ (see, e.g., \cite{Arnold:2008vd}) $C(q_\perp\gg T) = {2 C_A^2g^4n}/{q_\perp^4}$,
where $C_A=N_C$ is the number of colors, and $n$ is the number density per degree of freedom.
We find that the kernel at later times is peaked at the Debye mass (squares), while at early times the peak is shifted to smaller momenta. In the right panel, we compare the nonequilibrium angular averaged collision kernel with the thermal kernel obtained for the same energy density. We find that at early times the nonequilibrium kernel is enhanced for small momentum transfer as compared to the thermal case and suppressed for large momentum transfer.

\section{Conclusions}
Our results for the collision kernel $C(\vb q_\perp, \tau)$ during the initial stages in heavy-ion collisions are consistent with the recent extraction of the jet quenching parameter $\hat q$ \cite{Boguslavski:2023alu}, both in anisotropy and magnitude. Compared to thermal equilibrium, we find that the probability of small-momentum transfer is enhanced, especially for momentum transfer along the beam axis, while large-momentum transfer is suppressed, indicating the enhancement of small-angle scatterings. Our extraction of the nonequilibrium collision kernel represents a significant step towards a more realistic modeling of jet quenching during the initial stages.

\bibliography{bib} 

\end{document}